\documentclass[final,3p,times,twocolumn]{elsarticle}
\usepackage{graphicx}
\usepackage{amsmath}
\usepackage{hyperref}
\usepackage[version=3]{mhchem}
\usepackage[usenames]{color}
\graphicspath{{Figures/}}
\renewcommand{\dj}{d\kern-0.4em\char"16\kern-0.1em}
\renewcommand{\DJ}{\raise0.3ex\hbox{-}\kern-0.36em D}

\usepackage{lineno}

\journal{Journal of ...}

\begin{document}
\begin{frontmatter}
\title{Single crystals of DPPH
grown from diethyl ether and carbon disulfide solutions
--- Crystal structures, IR, EPR and magnetization studies}

\author[Rudjer]{Dijana \v{Z}ili\'{c}\corref{cor1}}
\ead{dzilic@irb.hr} \cortext[cor1]{Corresponding author. Fax: +
385 1 4680 245.}

\author[PMF]{Damir Paji\'{c}}

\author[Rudjer]{Marijana Juri\'{c}}

\author[Rudjer]{Kre\v{s}imir Mol\v{c}anov}

\author[Rudjer]{Boris Rakvin}

\author[Rudjer]{Pavica Planini\'{c}}

\author[PMF]{Kre\v{s}o Zadro}

\address[Rudjer]{Ru\dj er Bo\v{s}kovi\'{c} Institute,
\\Bijeni\v{c}ka cesta 54, 10000 Zagreb, Croatia}

\address[PMF]{Department of Physics, Faculty of Science,
University of Zagreb, \\Bijeni\v{c}ka cesta 32, 10000 Zagreb,
Croatia}

\begin{abstract}

Single crystals of the free radical 2,2-diphenyl-1-picrylhydrazyl
(DPPH) obtained from diethyl ether (ether) and carbon disulfide
(\ce{CS2}) were characterized by the X-ray diffraction, IR, EPR
and SQUID magnetization techniques. The X-ray structural analysis
and IR spectra showed that the DPPH form crystallized from ether
(DPPH1) is solvent free, whereas that one obtained from \ce{CS2}
(DPPH2) is a solvate of the composition
\mbox{4\,DPPH\,$\cdot$\,\ce{CS2}}. Principal values of the
$g$-tensor were estimated by the X-band EPR spectrometer at room
and low (10~K) temperatures. Magnetization studies revealed the
presence of antiferromagnetically coupled dimers in both types of
crystals. However, the way of dimerization as well as the strength
of exchange couplings are different in the two DPPH samples, which
is in accord with their crystal structures. The obtained results
improved parameters accuracy and enabled better understanding of
properties of DPPH as a standard sample in the EPR spectrometry.

\end{abstract}
\begin{keyword}
DPPH \sep crystal structure \sep diethyl ether \sep carbon
disulfide \sep EPR \sep magnetization



\end{keyword}

\end{frontmatter}

\linenumbers
\section{Introduction}

The stable aromatic free radical 2,2-diphenyl-1-picrylhydrazyl
(DPPH) is one of the first and most widely used standard samples
for determination of the $g$-factors of the spin species and for
measuring the unpaired spin concentration using electron
paramagnetic resonance (EPR) \cite{Krzystek1997}. DPPH was
synthesized in 1922, and its EPR spectrum was recorded for the
first time in 1950 \cite{Yordanov1996}. Chemical stability of DPPH
and its very narrow spectral line have led to the widespread use
of the powder form of this radical as an EPR standard
\cite{Boucherle1987}. Single crystals of DPPH are also frequently
used in EPR spectroscopy because the linewidth they produce is
considerably narrower than that of the powder form.

Various types of DPPH crystals have been prepared up to now ---
some of them are solvent free and some contain molecules of
solvation \cite{Weil1965}. In the Cambridge Structural Database
\cite{Allen2002} crystal structures of two DPPH solvates, one with
acetone \cite{Kiers1976} and the other with benzene
\cite{Williams1967}, are deposited. The benzene solvate was also
investigated by neutron diffraction \cite{Boucherle1987}. Some
preliminary X-ray diffraction measurements, done by Williams
\cite{Williams1965}, indicated  that the DPPH crystal forms
obtained from diethyl ether (ether; orthorhombic crystal system)
and carbon disulfide (\ce{CS2}; triclinic crystal system) were
both solvent free. However, their crystal structures have never
been solved, i.e.\ no atomic coordinates have been deposited in
the Cambridge Structural Database \cite{Allen2002}.

More recently, a new application of DPPH --- in detecting local
fields in the close vicinity of the surface of superconductors
\cite{Rakvin1989,Rakvin1990} and single molecule magnets
\cite{Rakvin2004} --- has been established in our laboratory.
These results prompted us to investigate the properties of DPPH in
more details.

In this paper, we report on the single-crystal X-ray diffraction
study, as well as the IR, EPR and SQUID magnetization measurements
of the two Williams' ''solvent-free'' forms \cite{Williams1965} of
DPPH, i.e. the one grown from ether (DPPH1) and the other form
crystallized from \ce{CS2} (DPPH2). A detailed structural analysis
showed that the orthorhombic DPPH form (crystallized from ether),
in accord with the previous preliminary measurements
\cite{Williams1965}, does not really contain solvent molecules;
however, the triclinic DPPH form (crystallized from \ce{CS2}),
believed to be solvent free for 40 years, is actually a solvate
with the stoichiometry \mbox{4\,DPPH\,$\cdot$\,\ce{CS2}}. This
form is isostructural with the acetone solvate,
\mbox{4\,DPPH\,$\cdot$\,\ce{CH3COCH3}} \cite{Kiers1976}. In
addition, it has been shown that magnetic properties of these two
kinds of DPPH crystals are quite different.

\section{Material and methods}

\subsection{Materials}
DPPH was purchased from commercial sources and used without
further purification. Elemental analysis for C, H and N was
carried out using a Perkin Elmer Model 2400 microanalytical
analyzer.

\subsection{Preparation of the single crystals}
\textbf{DPPH1.} Crystals of DPPH1 were grown from a solution of
DPPH in ether. The tightly closed reaction beaker was kept in a
refrigerator. The dark needle-like crystals were obtained after
two days. Anal.\,calcd for \ce{C18H12N5O6} ($M_r = 394.33$): C,
54.83; H, 3.07; N, 17.76. Found: C, 54.48; H, 3.32; N, 17.62\%. IR
data (KBr): $\tilde{\nu}$  = 3085 (w), 3071 (vw), 1598 (s), 1575
(s), 1523 (s), 1479 (m), 1460 (w), 1453 (w), 1434 (w), 1408 (w),
1324 (vs), 1292 (sh), 1212 (s), 1171 (m), 1073 (s), 1024 (w), 997
(w), 952 (m), 935 (w), 914 (m), 908 (sh), 842 (w), 833 (sh), 819
(w), 787 (m), 755 (s), 740 (m), 724 (m), 712 (m), 703 (m), 698
(m), 686 (s), 653 (w), 620 (w), 578 (w), 557 (w), 523 (w), 509
(w), 462 (w), 440 (w), 420 (w), 371 (w), 308 (w) cm$^{-1}$.

\textbf{DPPH2.} Crystals of DPPH2 were grown from a solution of
DPPH in \ce{CS2}. The tightly closed reaction beaker was kept in a
refrigerator. The dark needle-like crystals were formed in a
period of six days. Anal.\,calcd for
\ce{C18H12N5O6}$\cdot$0.25\ce{CS2} ($M_r = 413.36$): C, 53.03; H,
2.93; N, 16.94. Found: C, 52.78; H, 3.12; N, 16.79\%. IR data
(KBr): $\tilde{\nu}$  = 3087 (w), 3069 (vw), 1597 (s), 1574 (s),
1539 (m), 1525 (m), 1512 (s), 1478 (m), 1462 (w), 1453 (w), 1439
(w), 1412 (w), 1326 (vs), 1292 (m), 1210 (m), 1171 (m), 1073 (s),
1025 (w), 996 (w), 951 (m), 936 (w), 914 (sh), 909 (m), 844 (sh),
832 (w), 819 (w), 787 (w), 765 (sh), 757 (s), 739 (m), 715 (s),
703 (s), 698 (sh), 688 (m), 680 (m), 646 (w), 616 (w), 581 (w),
560 (w), 507 (w), 460     (w), 434 (w), 425 (w), 359 (w), 305 (w)
cm$^{-1}$.

\subsection{Physical techniques}

\textbf{Crystallography.} Single crystals of DPPH1 and DPPH2 were
measured on an Oxford Diffraction Xcalibur Nova diffractometer
with a microfocus copper tube (Cu$K_{\alpha}$ radiation) at room
temperature ($T=293\,(2)$~K). Lowering the temperature drastically
increased mosaicity, significantly degrading data quality.

CrysAlis PRO \cite{CrysAlis} program package was used for data
reduction. The structures were solved with SHELXS97 and refined
with SHELXL97 \cite{Sheldrick2008}. The models were refined using
the full-matrix least-squares refinement. All atoms except
hydrogen were refined anisotropically; hydrogen atoms were located
from the difference Fourier map and refined as riding entities.
The atomic scattering factors were those included in SHELXL97
\cite{Sheldrick2008}. Molecular geometry calculations were
performed with PLATON \cite{platon}, and molecular graphics were
prepared using ORTEP-3 \cite{Farrugia1997} and CCDC-Mercury
\cite{Mercury}. Crystallographic and refinement data for the
structures reported are shown in Table~\ref{tehnikalije}.

Supplementary crystallographic data for this paper can be obtained
free of charge via \\
www.ccdc.cam.ac.uk/conts/retrieving.html (or from the Cambridge
Crystallographic Data Centre, 12, Union Road, Cambridge CB2 1EZ,
UK; fax: +44 1223 336033; or deposit@ccdc.cam.ac.uk). CCDC 732147
\& 732148 contain the supplementary crystallographic data for this
paper.

\begin{table*}
\caption{Crystallographic, data collection and structure
refinement data.} \label{tehnikalije}

\centering \resizebox{10cm}{!} {\scriptsize{\begin{tabular}{l l l}

                                &                     &\\
\hline
                                &DPPH1                &DPPH2\\
\hline
Chemical formula                &\ce{C18H12N5O6}      &\ce{C18H12N5O6}$\cdot$0.25\ce{CS2}\\
\emph{M$_{r}$} / g mol$^{-1}$   &394.33               &413.36\\
                                &                     &\\
Color                          &black                &black\\
Crystal size / mm               &0.25\,x\,0.10\,x\,0.07   &0.28\,x\,0.13\,x\,0.08\\
Crystal system                  &orthorhombic         &triclinic\\
Space group                     &$P\,n\,2_{1}\,a$     &$P\,\overline{1}$\\
\emph{a} / \AA                  &16.7608\,(7)         &7.5577\,(5)\\
\emph{b} / \AA                  &26.8351\,(9)         &13.5724\,(7)\\
\emph{c} / \AA                  &7.8458\,(3)          &18.922\,(1)\\
$\alpha$ / $^{\circ}$           &90                   &95.084\,(4)\\
$\beta$ / $^{\circ}$            &90                   &92.141\,(5)\\
$\gamma$ / $^{\circ}$           &90                   &101.488\,(5)\\
\emph{V} / \AA$^3$              &3528.9\,(2)          &1891.6\,(2)\\
\emph{Z}                        &8                    &4\\
$D_{calc}$ / g\,cm$^{-3}$       &1.484                &1.451\\
                                &                     &\\
Radiation                       &Cu$K_{\alpha}$       &Cu$K_{\alpha}$\\
Data collection method          &CCD                  &CCD\\
\emph{T} / K                    &293\,(2)             &293\,(2)\\
Absorption correction           &none                 &none\\
Measured reflections            &11143                &19986\\
Independent reflections         &3664                 &7590\\
Observed reflections ($I>2\sigma(I)$) &2787           &3908\\
$R_{int}$                       &0.0387               &0.0545\\
$\Theta_{max}$ / $^{\circ}$         &76.29                &76.15\\
                                &                     &\\
Refinement                      &$F^{2}$              &$F^{2}$\\
$R[F^{2}>2\sigma F^{2}]$        &0.0674               &0.0639\\
$wR(F^{2})$                     &0.1746               &0.2151\\
\emph{S}                        &1.069                &0.963\\
No.\ of reflections             &3664                 &7590\\
No.\ of parameters              &523                  &538\\
H-atom treatment                &constrained          &constrained\\
$\Delta \rho_{max}$,            $\Delta \rho_{min}    $&0.308; -0.210&0.438; -0.391\\
\hline

\end{tabular}}}
\end{table*}

\textbf{IR spectroscopy.} Infrared spectra were recorded as KBr
pellets on an ABB Bomem FT model MB 102 spectrometer, in the
\mbox{4000--200~cm$^{-1}$} region.

\textbf{EPR spectroscopy.} EPR measurements were performed on the
single crystals of DPPH1 and DPPH2. Dimensions of the prepared
single crystals were approximately \mbox{$2.0 \times 0.2 \times
0.2$~mm$^3$}. The crystals were mounted on a quartz holder in the
cavity of an X-band EPR spectrometer (Bruker Elexsys 580 FT/CW)
equipped with a standard Oxford Instruments model DTC2 temperature
controller. The measurements were performed at the microwave
frequency around 9.7~GHz with the magnetic field modulation
amplitude of 5~$\mu$T at 100~kHz. The crystals were rotated round
three mutually orthogonal axes: a crystallographic $a$~axis (the
crystals of both DPPH1 and DPPH2 were elongated along the
$a$~axes), an arbitrary chosen $b^{\ast}$ axis perpendicular to
$a$ and a third $c^{\ast}$~axis, perpendicular to both $a$ and
$b^{\ast}$ (because of the thin needle-like form, it was difficult
to orientate crystals in the crystallographic $b$ and $c$ axes).
The EPR spectra were recorded at $5^{\circ}$ steps. The rotation
was controlled by a goniometer with the accuracy of
1--2$^{\circ}$. A larger uncertainty (2--3$^{\circ}$) was related
to the optimal deposition of the crystals on the quartz holder.
The EPR spectra were measured at two temperatures: room
\mbox{($T=297$~K)} and low \mbox{($T=10$~K)}.

\textbf{Magnetization study.} Magnetization of the DPPH1 and DPPH2
samples in the powdered form (about 25~mg) was measured using a
commercial MPMS5 SQUID magnetometer. The magnetization was checked
to be linear with respect to the applied magnetic field up to 5~T
for both compounds at several temperatures (2, 5 and 50~K). The
temperature dependence of magnetization was measured in the
applied magnetic fields of 0.1 and 1~T, in the temperature range
1.9--290~K. For each particular compound, measurements in the two
different magnetic fields resulted with identical susceptibility
\textit{vs} temperature curves.


\section{Results and discussion}

\subsection{Crystallography}
\begin{figure}
\centerline{\includegraphics[clip=,width=8cm]{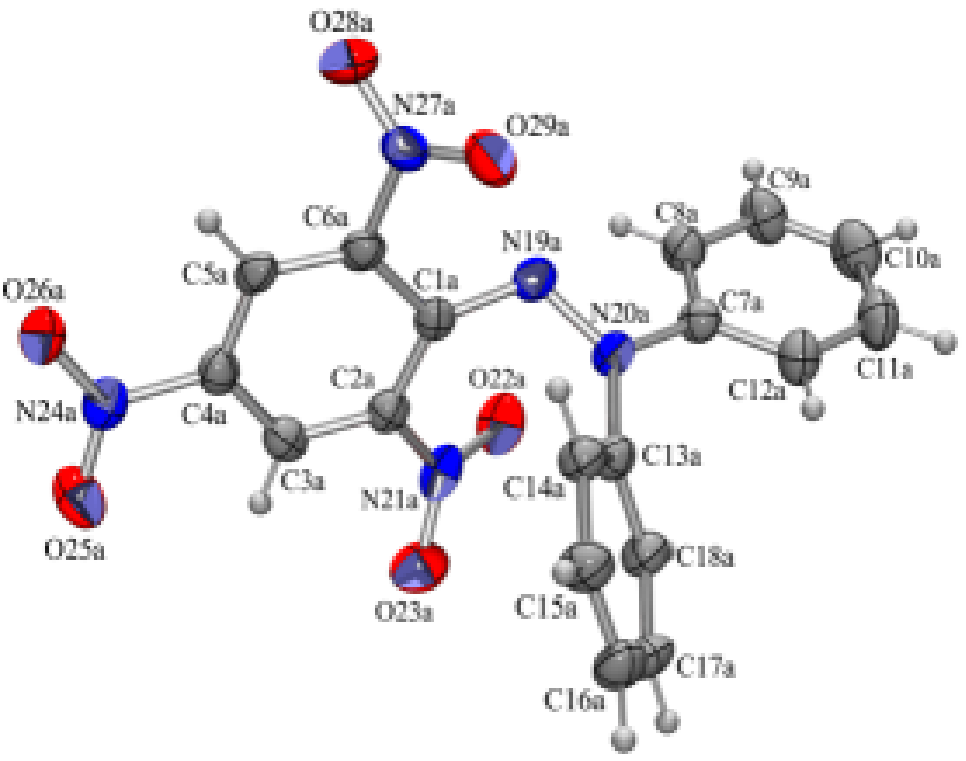}}
\centerline{\includegraphics[clip=,width=8cm]{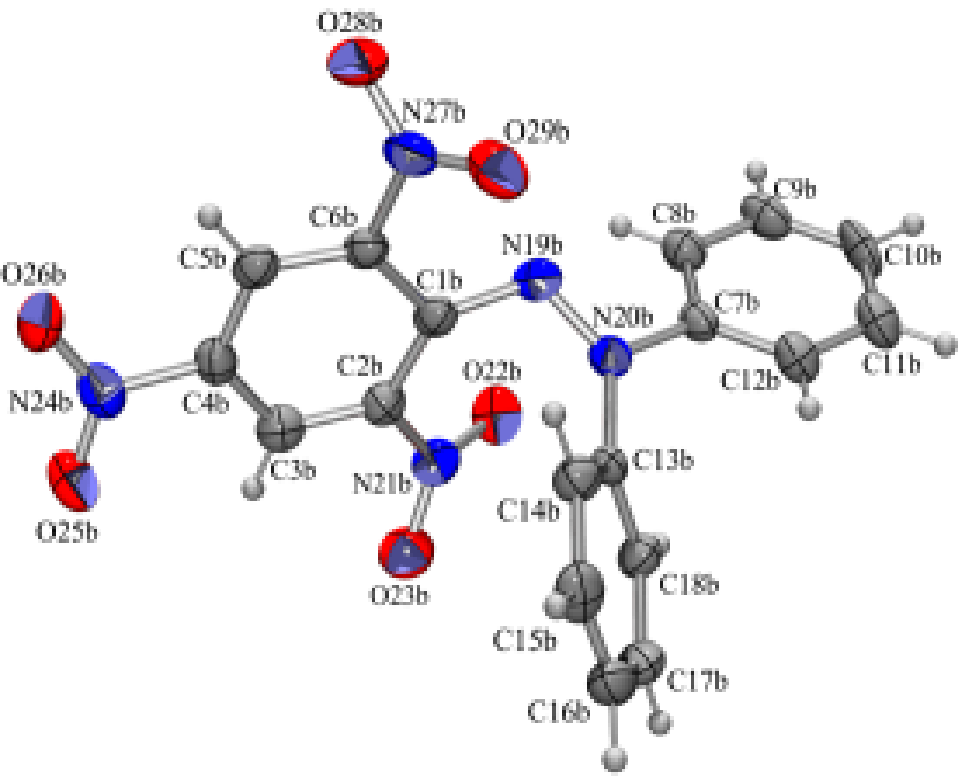}}
\caption{ORTEP-3 \cite{Farrugia1997} drawing of two
symmetry-independent molecules in DPPH1. Atomic displacement
ellipsoids are drawn at 50\% probability and hydrogen atoms are
depicted as spheres of arbitrary radii. Atom numbering is the same
as in other crystallographic studies
\cite{Williams1967,Kiers1976,Boucherle1987}; labels \textbf{a} and
\textbf{b} denote symmetry-independent molecules \textbf{a} and
\textbf{b}.} \label{DPPH1-ORTEP}
\end{figure}

\begin{figure}
\centerline{\includegraphics[clip=,width=8cm]{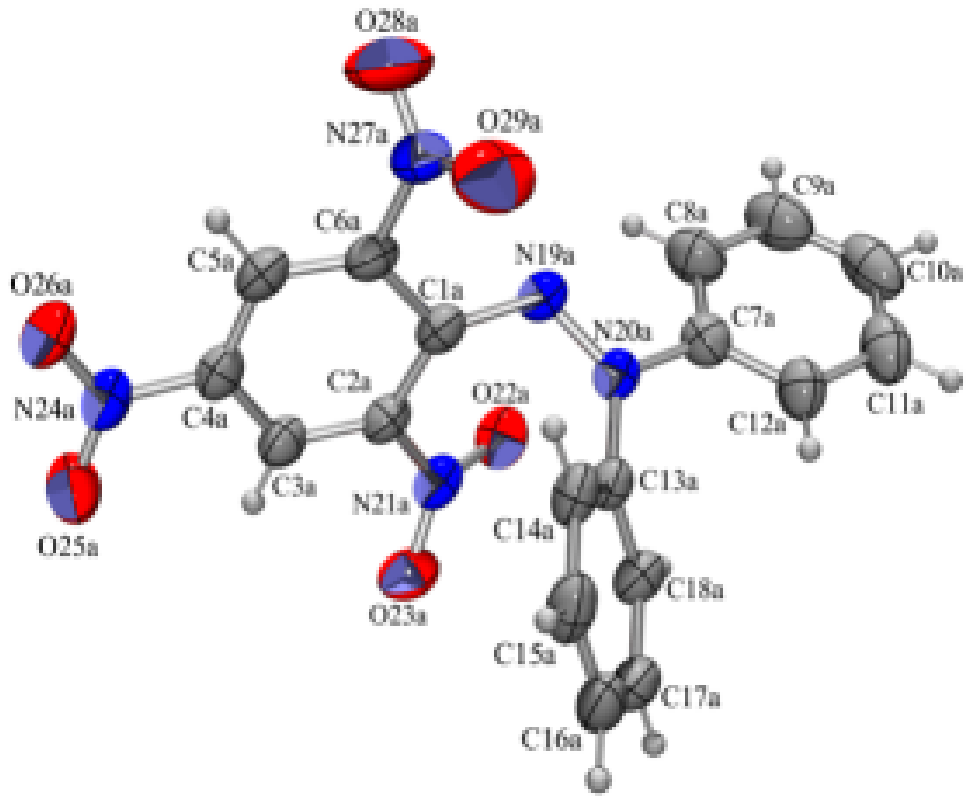}}
\centerline{\includegraphics[clip=,width=8cm]{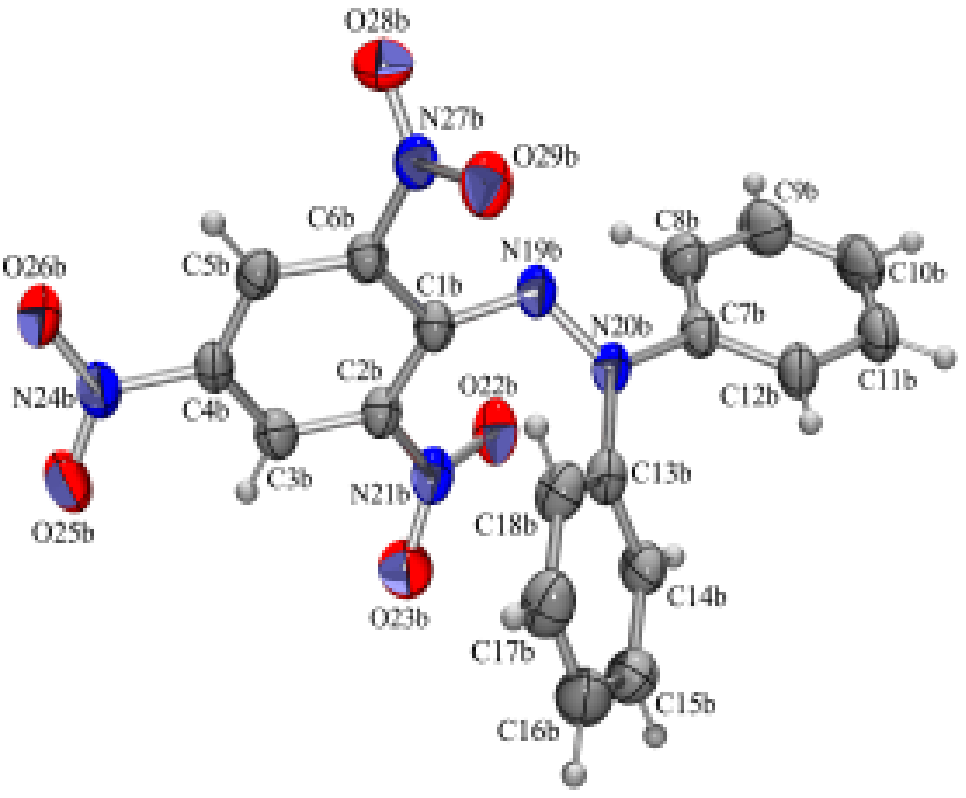}}
\caption{ORTEP-3 \cite{Farrugia1997} drawing of two
symmetry-independent molecules in DPPH2. Atomic displacement
ellipsoids are drawn at 50\% probability and hydrogen atoms are
depicted as spheres of arbitrary radii. Atom numbering is the same
as in other crystallographic studies
\cite{Boucherle1987,Kiers1976,Williams1967}; labels \textbf{a} and
\textbf{b} denote symmetry-independent molecules \textbf{a} and
\textbf{b}.} \label{DPPH2-ORTEP}
\end{figure}

The geometries and conformations of the DPPH radicals, DPPH1 and
DPPH2 (Figures~\ref{DPPH1-ORTEP} and \ref{DPPH2-ORTEP}), agree
well with those found in previous crystallographic studies of DPPH
solvates \cite{Boucherle1987,Kiers1976,Williams1967}. Bond lengths
and angles of the pycryl--N--N--Ph$_{2}$ system
(Table~\ref{NNgeom}) indicate that the unpaired electron is
delocalized over the \mbox{C1--N19--N20} fragment with the bonds
order of \emph{ca}~1.5. The bond order of \mbox{N20--C7} and
\mbox{N20--C13} is \emph{ca}~1. Such an electronic structure is in
agreement with a recent DFT study \cite{Mattar2006}. The DPPH
molecule is not rigid; however, ENDOR spectroscopy
\cite{Dalal1975} and DFT calculations \cite{Mattar2006} indicate
that restricted rotations of phenyl rings are possible in
solution. Therefore, the crystallographically observed
conformation is thermodynamically, probably, the most stable one.

\begin{table}
    \begin{center}
    \caption{Geometric parameters of the pycryl--N--N--Ph$_2$ system (\AA, $^{\circ}$).}
    \label{NNgeom}
    \resizebox{8cm}{!} {
    \scriptsize
  \begin{tabular}{l l l}
&&\\
\hline
        \multicolumn{2}{c}{\textbf{DPPH1}} \\
&molecule \textbf{a}&molecule \textbf{b}\\
C1--N19&1.364\,(8)&1.376\,(8)\\
N19--N20&1.352\,(7)&1.321\,(7)\\
N20--C7&1.405\,(8)&1.426\,(8)\\
N20--C13&1.432\,(8)&1.435\,(7)\\
C1--N19--N20&118.0\,(5)&117.0\,(5)\\
N19--N20--C7&116.9\,(5)&115.6\,(5)\\
N19--N20--C13&121.4\,(5)&123.5\,(5)\\
C7--N20--C13&121.0\,(5)&120.2\,(5)\\
        \multicolumn{2}{c}{\textbf{DPPH2}} \\
        &molecule \textbf{a}&molecule \textbf{b}\\
C1--N19&1.354\,(5)&1.366\,(4)\\
N19--N20&1.342\,(4)&1.339\,(4)\\
N20--C7&1.404\,(5)&1.416\,(4)\\
N20--C13&1.434\,(5)&1.432\,(4)\\
C1--N19--N20&\,118.6(3)&118.8\,(3)\\
N19--N20--C7&\,115.6(3)&115.6\,(3)\\
N19--N20--C13&\,122.0(3)&121.5\,(2)\\
C7--N20--C13&\,121.7(3)&122.5\,(2)\\
        \hline
        \end{tabular}}
        \end{center}
    \end{table}

\begin{figure}
\centerline{\includegraphics[clip=,width=7cm]{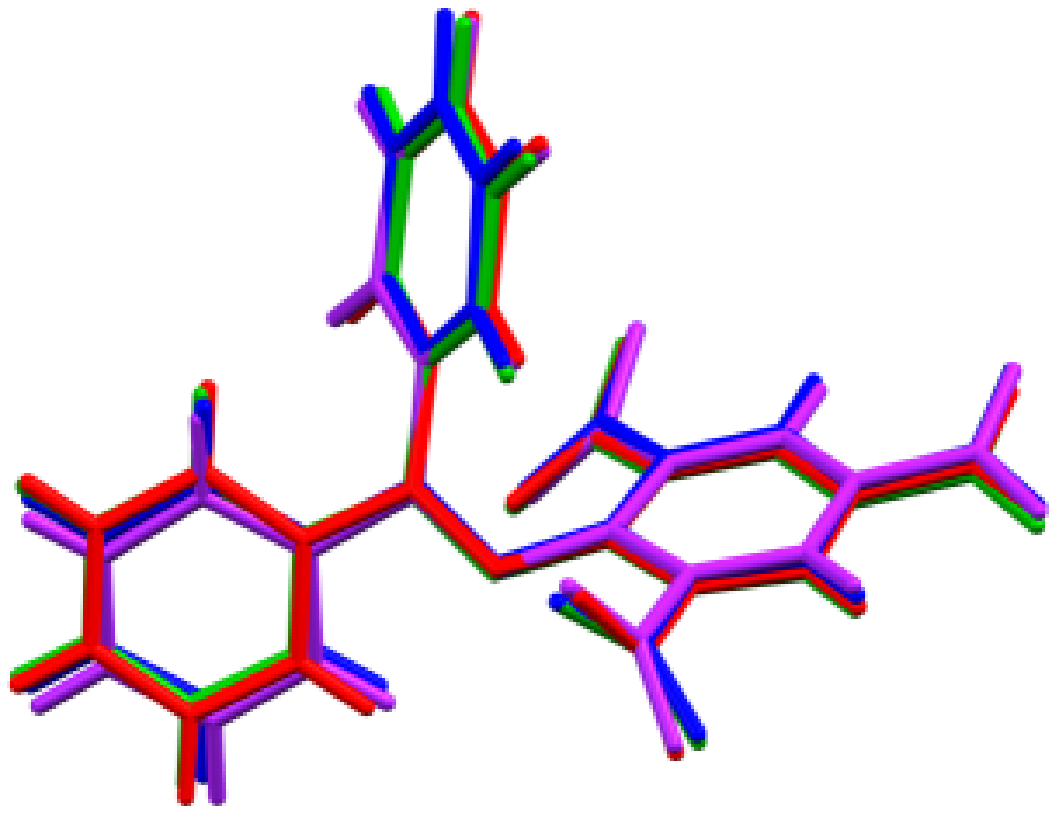}}
\caption{Overlap of four symmetry-independent molecules in the
crystal structures of DPPH1 and DPPH2. The differences in
conformations are almost all within 3 e.s.d.'s. Molecules
\textbf{a} and \textbf{b} of DPPH1 are green and blue, while
molecules \textbf{a} and \textbf{b} of DPPH2 are red and purple.}
\label{DPPH-overlej}
\end{figure}

\begin{figure}
\centerline{\includegraphics[clip=,width=8cm]{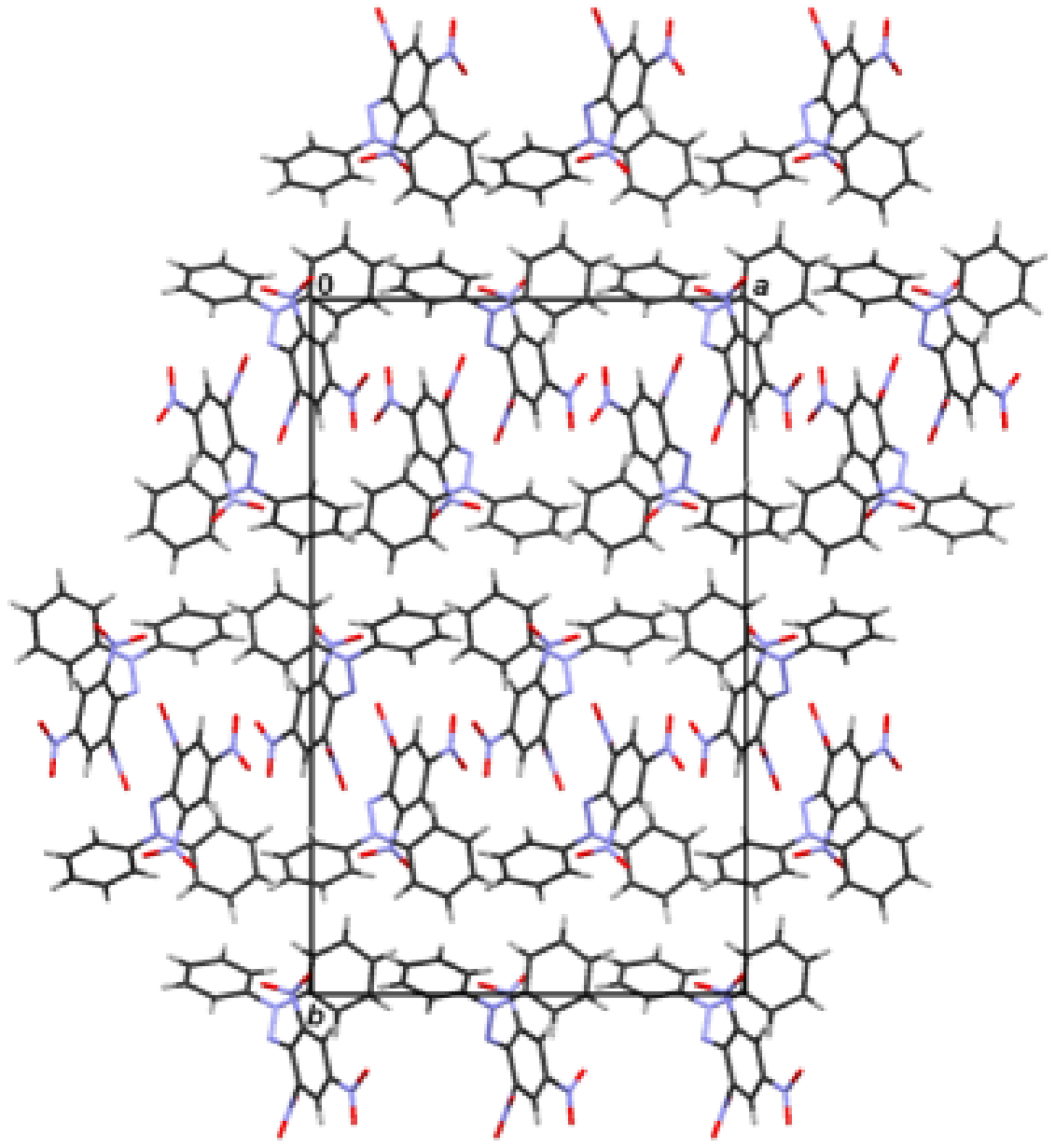}}
\caption{Crystal packing of DPPH1 viewed in the direction [001].
\ce{C-H\bond{...}O} hydrogen bonds have been omitted for clarity.}
\label{DPPH1-pak-001}
\end{figure}

\begin{figure}
\centerline{\includegraphics[clip=,width=8cm]{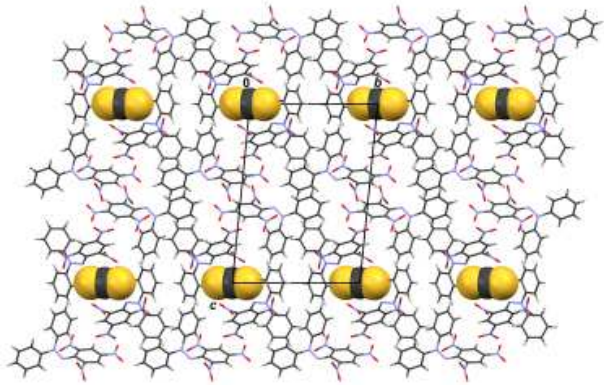}}
\caption{Crystal packing of DPPH2 showing channels containing
molecules of \ce{CS2} that run in the direction [100]. For
clarity, \ce{CS2} molecules are shown as van der Waals spheres.}
\label{DPPH2-pak-100}
\end{figure}

In the both DPPH1 and DPPH2 crystal structures, the asymmetric
unit contains two symmetry-independent DPPH radicals; the
asymmetric unit of DPPH2 contains also a half of a \ce{CS2}
molecule (its sulphur atom is located in a crystallographic
inversion center). All four symmetry-inequivalent molecules
described in the paper adopt the same conformation
(Figure~\ref{DPPH-overlej}), already observed in the crystal
structures of several DPPH crystal forms
\cite{Boucherle1987,Kiers1976,Williams1967}. Crystal packings of
the both structures (DPPH1 and DPPH2) are dominated by the
\ce{C\bond{-}H\bond{...}O} hydrogen bonds (Table~\ref{HB}). In
DPPH2, \ce{$\pi$\bond{...}$\pi$} interactions are also present
(Table~\ref{pipi}). DPPH1 forms a 3D hydrogen bonded network
(Figure~\ref{DPPH1-pak-001}), while DPPH2 forms 2D hydrogen bonded
sheets parallel with (100), held together by the
\ce{$\pi$\bond{...}$\pi$} interactions. Such a structure is
porous, with channels filled with \ce{CS2} molecules running in
the direction [100] (Figure~\ref{DPPH2-pak-100}).

\begin{table*}
    \begin{center}
    \caption{Geometric parameters of the hydrogen bonds (\AA, $^{\circ}$).}
    \label{HB}
    \resizebox{14cm}{!} {
    \scriptsize
  \begin{tabular}{l l l l l l}
        \hline
        &&&&&\\ & $d$(\emph{D}--H)\,/\,\AA& $d$(H$\cdot\cdot\cdot$\emph{A})\,/\,\AA& $d$(\emph{D}$\cdot\cdot\cdot$\emph{A})\,/\,\AA&(\emph{D}--H$\cdot\cdot\cdot$\emph{A})\,/\,$^{\circ}$&Symm.\ op.\\
        &&&&&\\
        \hline
        \multicolumn{2}{c}{\textbf{DPPH1}} \\
        C14A--H14A$\cdot\cdot\cdot$O25A&0.93&2.53&3.151\,(8)&125&$x, y, -1+z$\\
        C14B--H14B$\cdot\cdot\cdot$O25B&0.93&2.54&3.185\,(8)&127&$x, y, -1+z$\\
      C17B--H17B$\cdot\cdot\cdot$O22A&0.93&2.53&3.400\,(10)&156&$1-x, 1/2+y, 1-z$\\
      C12B--H12B$\cdot\cdot\cdot$O23B&0.93&2.64&3.344\,(9)&145&$x, y, 1-+z$\\
      C8A--H8A$\cdot\cdot\cdot$O26B&0.93&2.61&3.278\,(4)&130&$x, y, -1+z$\\
        \multicolumn{2}{c}{\textbf{DPPH2}} \\
        C14A--H14A$\cdot\cdot\cdot$O25A&0.93&2.61&3.355\,(6)&135&$-1+x, y, z$\\
        C12A--H12A$\cdot\cdot\cdot$O23A&0.93&2.63&3.301\,(5)&129&$-1+x, y, z$\\
        C5B--H5B$\cdot\cdot\cdot$O28B&0.93&2.72&3.355\,(6)&127&$1-x, 2-y, 1-z$\\
        C15A--H15A$\cdot\cdot\cdot$O26B&0.93&2.69&3.493\,(6)&145&$-1+x, -1+y, z$\\
        C8A--H8A$\cdot\cdot\cdot$O28A&0.93&2.54&3.200\,(7)&129&$-x, -y, -z$\\
        \hline
        \end{tabular}}
        \end{center}
    \end{table*}

    \begin{table*}
    \begin{center}
    \caption{Geometric parameters of \ce{$\pi$\bond{...}$\pi$} interactions in DPPH2 (\AA, $^{\circ}$).}
    \label{pipi}
    \resizebox{14cm}{!} {
    \scriptsize{
  \begin{tabular}{l l l l l l l}
        \hline
        & $Cg$$^{1}\cdot\cdot\cdot$ $Cg$ &$\alpha^{2}$&$\beta^{3}$&$\delta^{4}$&offset\,/\,\AA&symm.op.\\
        \hline
        C1B\ce{->}C6B$\cdot\cdot\cdot$C1B\ce{->}C6B&4.027\,(2)&0.00&32.17&3.409&2.144&$2-x, 2-y, 1-z$\\
        C7B\ce{->}C12B$\cdot\cdot\cdot$C7B\ce{->}C12B&3.973\,(1)&0.00&20.16&3.730&1.369&$1-x, 1-y, 1-z$\\
        \hline
        \end{tabular}}
        }
                \begin{flushleft}\scriptsize{
        $^{1}$Ring centroid;
        $^{2}$Angle between two ring planes;
        $^{3}$Angle between a centroid-centroid line and a normal to the plane of the first ring;
        $^{4}$Distance between the centroid of the first ring and the plane of the second
        one.}
        \end{flushleft}
        \end{center}
        \end{table*}


\subsection{IR spectroscopy}
The IR spectra of DPPH1 and DPPH2 show characteristic absorption
bands that can, in general, be attributed to the presence of
aromatic hydrocarbon ligands and nitro groups. The absorption
bands of weak intensity that occur in the region
3100--3000~cm$^{-1}$ for both compounds originate from the
aromatic \mbox{C--H} stretching vibrations. The absorption bands
of rather strong intensity at 1598, 1575 and 1479~cm$^{-1}$ in the
spectrum of DPPH1, and at 1597, 1574 and 1478~cm$^{-1}$ in the
spectrum of DPPH2 correspond to the stretch of the \mbox{C--C}
bonds from the aromatic rings \cite{Silverstein1974}. The
absorption bands corresponding to the nitro groups in DPPH1 are
located at 1523~cm$^{-1}$ [$\nu_{as}$(NO)], 1324~cm$^{-1}$
[$\nu_{s}$(NO)], 914~cm$^{-1}$ [$\nu$(C--NO$_2$)] and also at 842
and 833~cm$^{-1}$ [$\delta$(ONO)]. The corresponding bands for
DPPH2 are placed at 1525~cm$^{-1}$ [$\nu_{as}$(NO)],
1326~cm$^{-1}$ [$\nu_{s}$(NO)], 914~cm$^{-1}$ [$\nu$(C--NO$_2$)]
and at 844 and 832~cm$^{-1}$ [$\delta$(ONO)]
\cite{Silverstein1974,Nakamoto1997}. The presence of the solvate
molecule of \ce{CS2} in DPPH2 is confirmed by the strong
absorption band at 1512~cm$^{-1}$ [$\nu_{as}$(\ce{CS2})]
\cite{Nakamoto1997}.

\subsection{EPR spectroscopy}

\subsubsection{DPPH1}
The EPR spectrum of DPPH1 was a Lorentzian singlet line at room
temperature. The angular rotation of the single crystal gave an
approximately isotropic line with the (peak-to-peak) width
\mbox{$W=(0.16\pm0.02)$~mT} and \mbox{$g=2.0036\pm0.0001$}.

\

\begin{figure}
\centerline{\includegraphics[width=6cm,clip=]{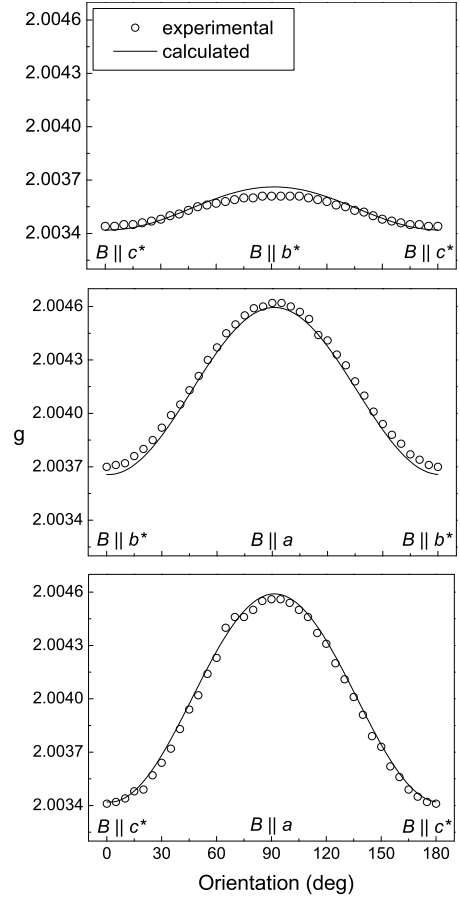}}
\caption{Angular variation of the $g$-values of EPR lines of the
single crystal of DPPH1 at $T=10$~K in three mutually
perpendicular planes. Experimental values are given by circles
 and solid lines represent calculated $g$-values.} \label{DPPH1}
\end{figure}

The temperature dependence of the linewidth was examined in the
range \mbox{$T=10$--297~K}. The results showed no significant
changes of this parameter. The singlet line at $T=10$~K had
approximately the same width \mbox{$W=(0.14\pm0.01)$~mT} as the
line at $T=297$~K. However, in contrary to the isotropic spectral
line at room temperature, the measurements at 10~K showed the
anisotropy of the spectrum. The angular variations of the
$g$-value of the single crystal rotated along the three chosen
orthogonal axes: $a$, $b^{*}$ and $c^{*}$ are shown in
Figure~\ref{DPPH1}.

The elements of the \mbox{$(\mathbf{g^{T}g})_{ij}$} matrix at
$T=10$~K were determined from the experimental single-crystal
data, by solving the following equation \cite{Weil1994}:
\begin{eqnarray}\label{gg}
 g^{2}=&(\mathbf{g^{T}g)}_{aa}\sin^{2}\theta\cos^{2}\phi +
         (\mathbf{g^{T}g)}_{ab}\sin^{2}\theta\sin2\phi +{} \nonumber\\
     &{}+(\mathbf{g^{T}g})_{bb}\sin^{2}\theta\sin^{2}\phi +
         (\mathbf{g^{T}g)}_{ac}\sin2\theta\cos\phi + {}\nonumber\\
     &{}+(\mathbf{g^{T}g)}_{bc}\sin2\theta\sin\phi +
         (\mathbf{g^{T}g)}_{cc}\cos^{2}\theta
\end{eqnarray}
where $\theta$ and $\phi$ are the polar and azimuthal angles of
the magnetic field vector $\mathbf{B}$ in the $a-b^{*}-c^{*}$
coordinate system, respectively. The calculated $g$-tensor is
presented in Figure~\ref{DPPH1} by solid lines. The principal
values of the \mbox{$g$-tensor} of DPPH1, obtained by
diagonalization of the \mbox{$\mathbf{g^{T}g}$} matrix at
\mbox{$T=10$~K}, are shown in Table~\ref{tab1}, with the estimated
error \mbox{$\pm$ 0.0001}. The obtained \mbox{$g$-tensor} is
approximately axial with the maximum value,
\mbox{$g_{xx}=2.0046$}, observed in the direction roughly parallel
to the crystallographic $a$~axis.

\begin{table}
\begin{center}
\caption{Principal values of the $g$-tensors of DPPH1 and DPPH2.}
\label{tab1}
\begin{tabular}{lccccc}
                                   &         &              &               & \\
\hline
                                   &  $T$ (K)& $g_{xx}$     & $g_{yy}$      & $g_{zz}$ \\

 \hline
                           DPPH1   &  297    &  2.0036      &    2.0036     & 2.0036  \\
                                   &   10    &  2.0046      &    2.0037     & 2.0034  \\
 \hline
                           DPPH2   &  297    &  2.0041      &    2.0036     & 2.0030  \\
                                   &   10    &  2.0055      &    2.0040     & 2.0024  \\
         DPPH \cite{Chirkov1970}  &  297    &  2.0037      &    2.0036     & 2.0034  \\
\hline
\end{tabular}
\end{center}
\end{table}

\subsubsection{DPPH2}

\begin{figure}
\centerline{\includegraphics[width=6cm,clip=]{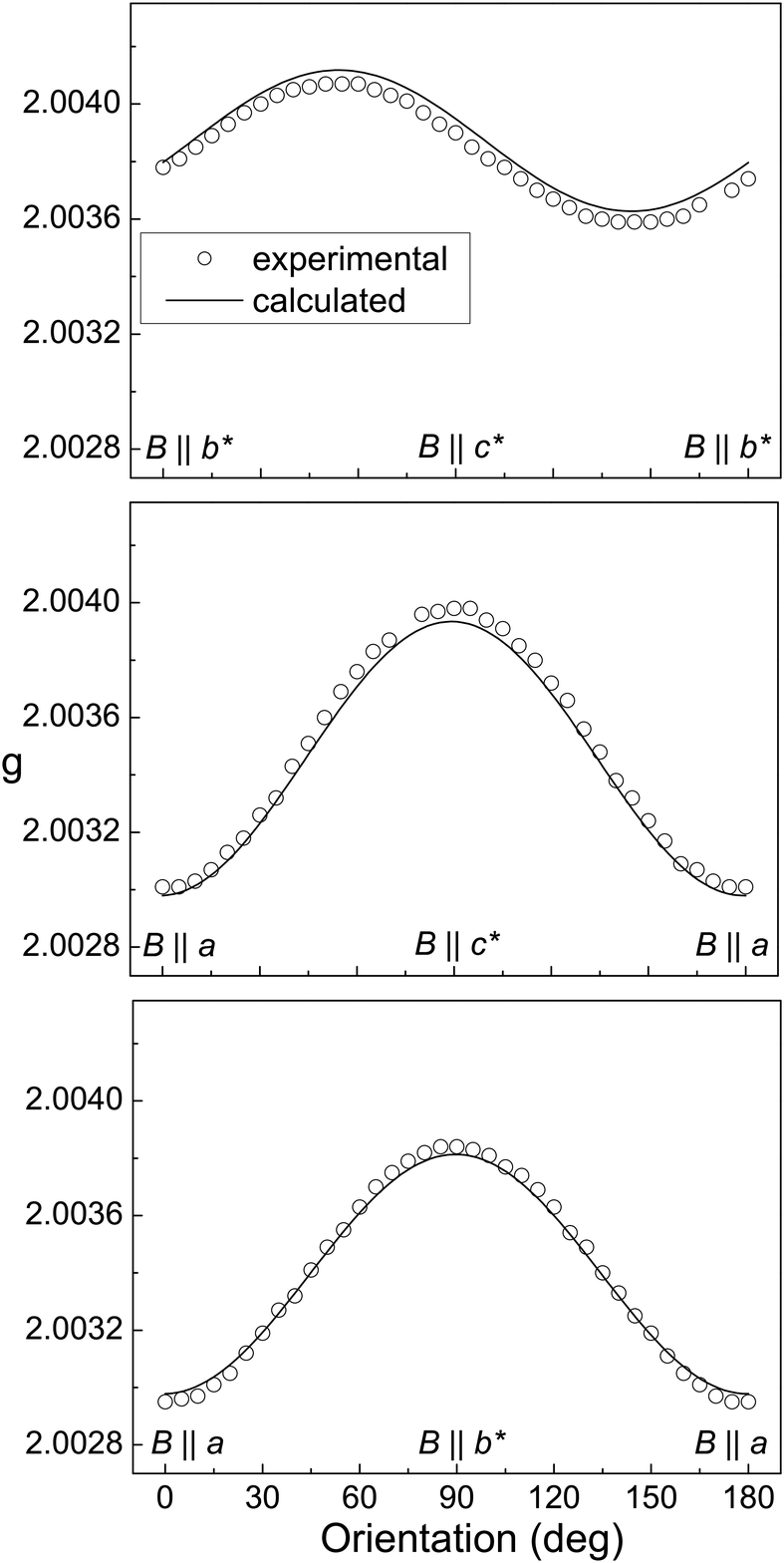}}
\caption{Angular variation of the $g$-values of EPR lines of the
single crystal of DPPH2 at $T=297$~K in three mutually
perpendicular planes. Experimental values are given by circles and
solid lines represent calculated $g$-values.} \label{DPPH2-297K}
\end{figure}

The single-crystals of DPPH2 showed the anisotropy singlet line
already at room temperature.  In comparison to DPPH1, the
linewidth of DPPH2 was almost halved: \mbox{$W=(0.08\pm0.02)$~mT}.
This is in agreement with the fact that the solid state EPR
spectrum of DPPH has a solvent dependent linewidth and  that the
lowest observed value of the linewidth was obtained for DPPH
crystallized from \ce{CS2} (0.15~mT for powder)
\cite{Yordanov1996}. Such a small value of the linewidth had
earlier led to the conclusion that the DPPH single crystals
obtained from \ce{CS2} were probably solvent free, although there
was no unambiguous evidence for that. The crystal structure data
presented in this study have undoubtedly showed that DPPH2 has
syncrystallized molecules of \ce{CS2}: the unit cell contains four
DPPH radicals and one \ce{CS2} molecule. The obtained linewidth,
in spite of the presence of the solvation, is very narrow
($\sim0.18$~mT for powder). The angular variations of the
$g$-value of DPPH2 along the three orthogonal axes at $T=297$~K
are shown in Figure~\ref{DPPH2-297K}. The dependencies obtained
are in approximate agreement with the earlier measurements that
had been performed round one axis for which crystallographic
indices had not been given \cite{Kikuchi1954,Singer1955}.

Using the same method as for DPPH1, the \mbox{$\mathbf{g^{T}g}$}
matrix was obtained and the principal values of the
\mbox{$g$-tensor} of DPPH2 at $T=297$~K were extracted and
presented in Table~\ref{tab1}. The minimum value of the
$g$-tensor, \mbox{$g_{zz}=2.0030$}, was observed in the direction
roughly parallel to the crystallographic $a$~axis.

The only up to now available experimental data for the $g$-tensor
of DPPH crystallized from \ce{CS2},  were obtained by Chirkov and
Matevosyan \cite{Chirkov1970}. They found different principal
values of the $g$-tensors for the crystals prepared under
different crystallization conditions (solvent purity, temperature,
\dots ) and explained this effect by the crystal lattice defects.
Otherwise, based on the fact that on raising the temperature right
up to the melting point the EPR linewidth altered smoothly, the
authors concluded that the crystals of DPPH had no solvent
(\ce{CS2}) molecules included. One set of the principal values of
the $g$-tensor obtained in the mentioned work is presented in
Table~\ref{tab1}, together with the corresponding values for DPPH1
and DPPH2. It could be seen that the $g$-tensor anisotropy
obtained by Chirkov and Matevosyan \cite{Chirkov1970} is
significantly lower than the anisotropy obtained in this study. A
reasonable explanation of the observed difference could be that
the crystals grown from different experimental conditions had
different CS2 : DPPH ratios and possibly, different crystal
structures.

\

\begin{figure}
\centerline{\includegraphics[width=6cm,clip=]{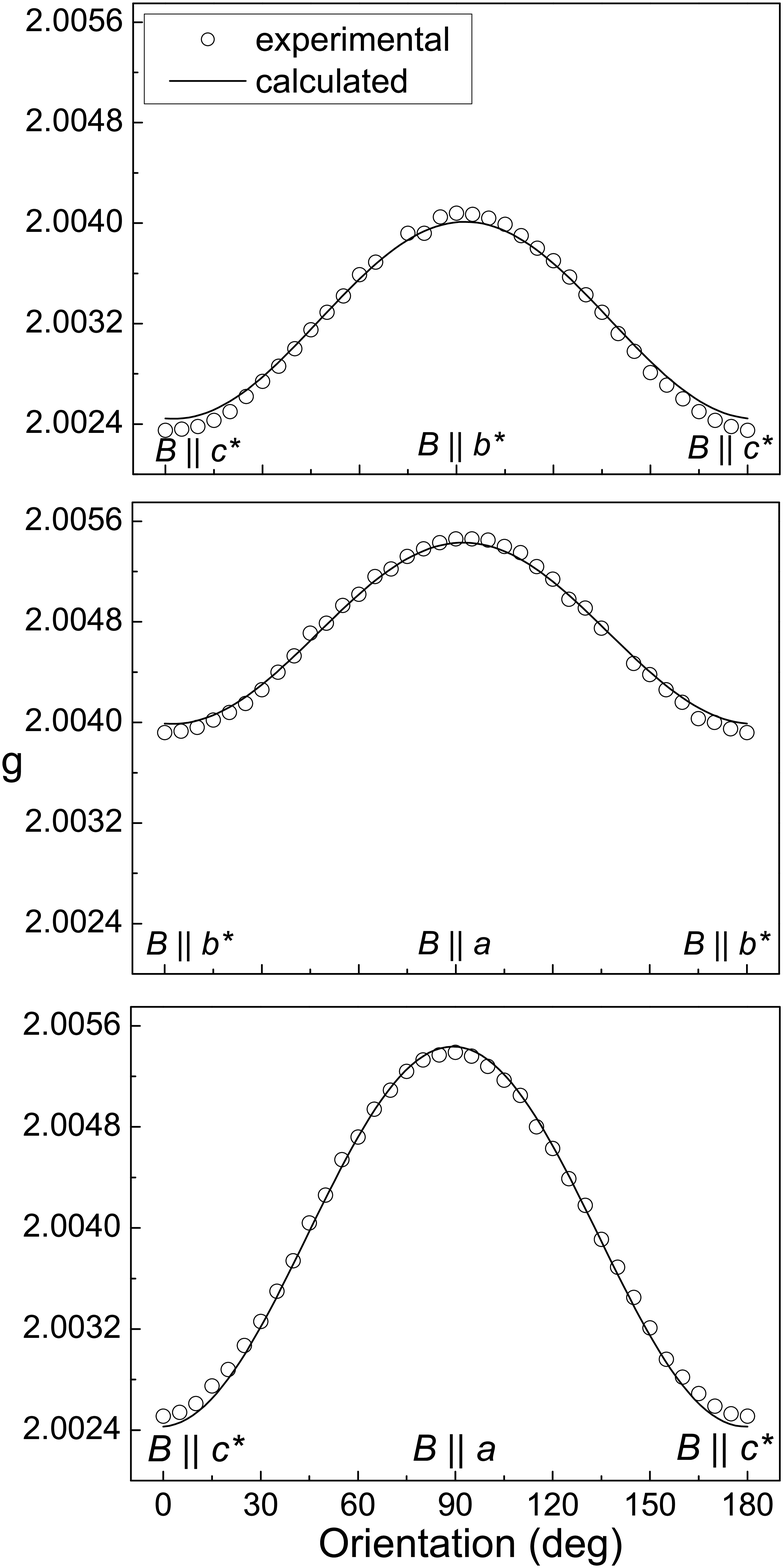}}
\caption{Angular variation of the $g$-values of EPR lines of the
single crystal of DPPH2 at $T=10$~K in three mutually
perpendicular planes. Experimental values are given by circles and
solid lines represent calculated $g$-values.} \label{DPPH2-10K}
\end{figure}

The linewidth of $W=(0.15\pm0.04)$~mT obtained for DPPH2 at
$T=10$~K shows a significant broadening compared to the linewidth
measured at room temperature. That is in agreement with the
earlier measurements for the powder DPPH form
\cite{Singer1955,Swarup1960}. The angular variations of the
\mbox{$g$-value} are shown in Figure~\ref{DPPH2-10K}.

The calculated principal values of the $g$-tensor of DPPH2 at
$T=10$~K are given in Table~\ref{tab1}. The obtained $g$-tensor
has the maximum value, \mbox{$g_{xx}=2.0055$}, in the direction
roughly parallel to the $a$~axis. Comparing
Figures~\ref{DPPH2-297K} and \ref{DPPH2-10K}, beside a change in
magnitude of the principal values of the $g$-tensor, also a shift
of the direction of eigenvectors could be observed. This effect
had been indicated earlier \cite{Singer1955}.


\subsection{Magnetization study}

\begin{figure}
\centerline{\includegraphics[width=8cm,clip=]{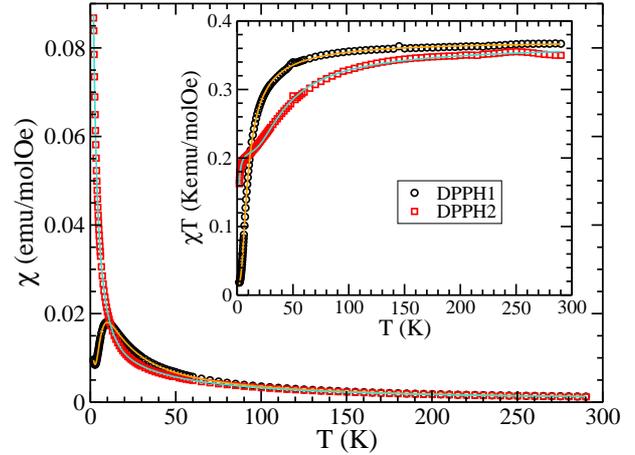}}
\caption{Temperature dependence of molar magnetic susceptibility
for DPPH1 (black circles) and DPPH2 (red squares) compounds. Solid
lines are the fitted curves. Inset: $\chi \cdot T$ \emph{vs} $T$
plot.} \label{magnetization}
\end{figure}

The temperature dependence of the molar magnetic susceptibility
$\chi$ for DPPH1 and DPPH2 is presented in Figure
\ref{magnetization}. The two DPPH samples show almost identical
behavior at temperatures above $\approx 150$~K, but their behavior
is qualitatively different at lower temperatures. For DPPH2, the
molar susceptibility $\chi $ is decreasing monotonously with
increasing temperature. For DPPH1, the susceptibility dependence
on temperature curve attains a relatively broad maximum at
\mbox{$T_{max}=10$~K}. The decrease of the $\chi $ value with
decreasing $T$ below $T_{max}$ points to the antiferromagnetic
interactions in this compound.

In the inset of Figure \ref{magnetization} the temperature
independent $\chi \cdot T$ value above 150~K was obtained after
the diamagnetic corrections of $-0.000180$ and $-0.000190$~emu/mol
for DPPH1 and DPPH2, respectively, were included. These values are
in agreement with those in the previously published work
\cite{Duffy1967}. From the $\chi \cdot T$ plots above 150~K the
Curie constant values of 0.363 and 0.351~emuK/mol for DPPH1 and
DPPH2, respectively, resulted. The values are close to the free
electron value of  0.375~emuK/mol, and according to the EPR
measurements, this should be the case also for DPPH1 and DPPH2.
From the EPR data, using $g=2.0036$, one can see that there are
96.5\% radical electrons of the spin $S=1/2$ per formula unit of
DPPH1 and 93.3\% radical electrons per formula unit of DPPH2.
Approximately the same values for the Curie constant $C$ result
from the Curie-Weiss analysis of $\chi ^{-1}(T)=(T-\theta )/C$,
where the slope of the straight line gives $C=0.373$~emuK/mol for
DPPH1 and $C=0.372$~emuK/mol for DPPH2. From here, the Curie-Weiss
parameter $\theta $ amounts $-5.3$ and $-13.8$~K for DPPH1 and
DPPH2, respectively. The negative values of this parameter point
to the antiferromagnetic interactions in both samples. The
obtained values are somewhat smaller than for other measured DPPH
crystals, where the $\theta $ values were found to be from $-22$
to $-26$~K \cite{vanItterbeek1964,Duffy1962}. However, it should
be noted that these parameters are empirical and descriptive only,
and might not give the true values of interaction energies.

\begin{figure}
\centerline{\includegraphics[width=7cm,clip=]{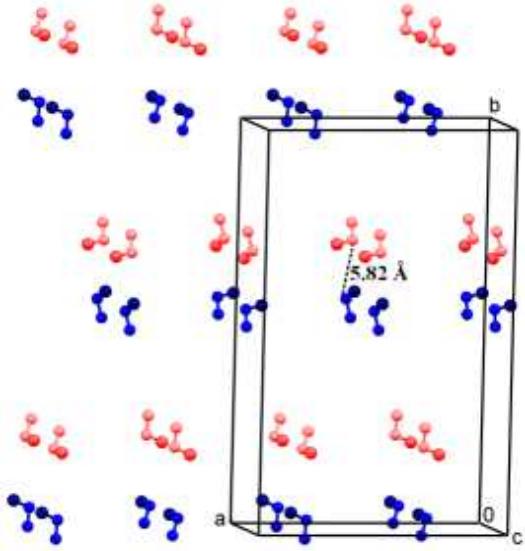}}
\caption{The closest distance between central N-atoms in the C-N-N
fragments in the crystal structure of DPPH1. The closest distance
is between two symmetry-independent molecules in the asymmetric
unit. Molecules are color-coded: those labelled as \textbf{a} are
red, lighter and \textbf{b} are blue, darker; carbon atoms are
drawn in a darker shade.} \label{DPPH1-interaction}
\end{figure}

\begin{figure}
\centerline{\includegraphics[width=8cm,clip=]{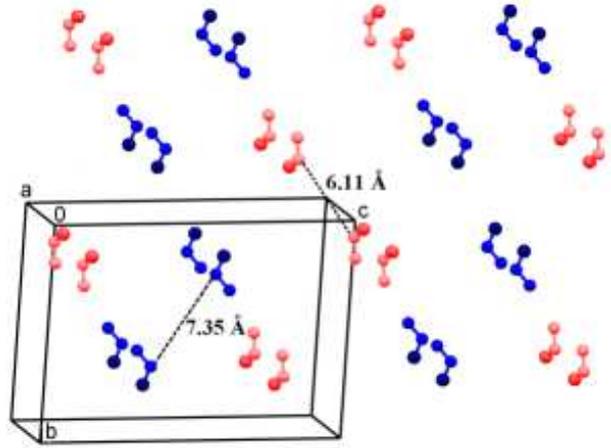}}
\caption{The closest distances between central N-atoms in the
C-N-N fragments in the crystal structure of DPPH2. The closest
distances are between pairs of molecules related by inversion
centers. Molecules are color-coded: those labelled as \textbf{a}
are red, lighter and \textbf{b} are blue, darker; carbon atoms are
drawn in a darker shade.} \label{DPPH2-interaction}
\end{figure}

The antiferromagnetic interactions for both compounds are
indicated by the downward bending of the $\chi \cdot T$ curves
with decreasing temperature (Figure \ref{magnetization}). Magnetic
correlations have visible effects starting approximately from 50
and 150~K for DPPH1 and DPPH2, respectively. Moreover, it seems
that for DPPH2 there are two characteristic temperatures
(energies). For further discussion of magnetic behavior of these
compounds, their structural characteristics should be taken into
account. It appears that consideration of the 3D long-range
interactions would not be appropriate, as no pathways for such
interactions could be observed in the crystal structures. Instead,
a more precise interpretation of the magnetic data should be found
in dimer interactions of radical electrons. Dimer approach was
reported earlier for other DPPH crystals
\cite{Duffy1967,Duffy1978}. Based on the crystal structures, such
an approach is also justified for the present DPPH samples.
Figures \ref{DPPH1-interaction} and \ref{DPPH2-interaction}
present simplified schemes of magnetic interactions in DPPH1 and
DPPH2 crystals, respectively. Only the C-N-N fragments are shown,
with the closest distances between the central N atoms (according
to the crystallographic and DFT studies, the unpaired electron is
delocalized over the \mbox{C1-N19-N20} bonds).

It is easy to notice that all molecules in DPPH1 are coupled into
dimers (Figure \ref{DPPH1-interaction}). Pairs of symmetry
independent molecules (in the same asymmetric unit) group into
dimers with the centroid distances of 5.82~{\AA}. Other
paramagnetic neighbors are mutually much more distant (more than
7~\AA).

In DPPH2 two kinds of dimers are observed (Figure
\ref{DPPH2-interaction}). In this structure, the closest
interactions are found between the pairs of molecules related by
the inversion centers. Those labelled as \textbf{a} (red, lighter
molecules) are mutually closer (6.11~\AA) than those labelled as
\textbf{b} (blue, darker molecules, 7.35~\AA). It could be
concluded that the DPPH2 molecules are divided into two types of
dimers with different distances between the unpaired electrons,
which could lead to different exchange parameters.

According to the previously mentioned, the susceptibility of DPPH1
is fitted by the following equation:
\begin{eqnarray}\label{chiDPPH1}
\chi (T) = w_{1}/2 \cdot \chi _{dim}(J) + w_{2} \cdot \chi _{CW},
\end{eqnarray}
where $w_{1}$ is the relative amount of molecules coupled into
dimers and $w_{2}$ is the relative amount of single molecules
interacting weakly with other neighboring molecules. The uncoupled
single paramagnetic centers could originate from the defects and
surface effects in the crystals. The susceptibility of dimers is
given by:
\begin{eqnarray}\label{chi_dim}
\chi _{dim}(J) = 2N\mu _{B}^{2}g^{2}/kT(3+\exp (-J/kT)),
\end{eqnarray}
where $J$ is the Heisenberg exchange coupling (defined by the
interaction Hamiltonian ${\cal H}_{INT}=-JS_{1}S_{2}$) between two
unpaired electrons in a dimer \cite{Kahn1993}. Other parameters
have their usual meanings. The Curie-Weiss molar susceptibility of
weakly interacting spin $S=1/2$ molecules is:
\begin{eqnarray}\label{chi_CW}
\chi _{CW} = N\mu _{B}^{2}g^{2}/4k(T-\theta ).
\end{eqnarray}
When $g=2.0036$ is assumed in accordance with the EPR
determination, the best fit is achieved with \mbox{$w_{1}=0.870$}
and \mbox{$w_{2}=0.0944$}. At the same time, the obtained
antiferromagnetic exchange coupling within the dimers is
$J=-17.5$~K and the long-range interaction Curie-Weiss parameter
$\theta =-1.89$~K. The agreement between the measured data and the
fitted function is excellent in the whole interval of temperature
(see Figure \ref{magnetization}). The value of $J$ is close to the
already published data on other DPPH crystals
\cite{Duffy1967,Duffy1978}.

The DPPH2 susceptibility was analyzed assuming the coexistence of
two kinds of dimers with different exchange couplings. Therefore,
the data were fitted by:
\begin{eqnarray}\label{chiDPPH2}
\chi (T) = w_{1}/2 \cdot \chi _{dim}(J_{1}) + w_{2}/2 \cdot \chi _{dim}(J_{2}),
\end{eqnarray}
where $ w_{1}$ and $ w_{2}$ are the relative amounts of molecules
coupled into particular types of magnetic dimers with the exchange
energies $J_{1}$ and $J_{2}$, respectively. The obtained
parameters are $w_{1}=0.570$ and $w_{2}=0.402$, whereas the
corresponding exchange interactions are $J_{1}=-1.56$~K and
$J_{2}=-83.9$~K. In Figure \ref{DPPH2-interaction}, $J_{1}$ and
$J_{2}$ could be associated with the molecules labelled as
\textbf{b} (blue, darker molecules) and \textbf{a} (red, lighter
molecules), respectively.

The fitting of the $\chi \cdot T$ curves (inset in Figure
\ref{magnetization}) gave consistently the same parameters for
both compounds. However, the obtained results for the exchange
parameters for DPPH1 ($J=-17.5$~K) and for the \textbf{a} labelled
molecules in DPPH2 (red, lighter molecules, $J_{2}=-83.9$~K),
which have approximately the same mutual distance within dimers,
are significantly different. The difference arises from the
different orientation of molecules in DPPH1 and DPPH2. The
molecules forming dimers in DPPH1 are almost mutually
perpendicular (the angle between the planes which are determined
by the C-N-N fragment, is 80.4$^{\circ}$) and the molecules
forming dimers in DPPH2 are mutually parallel (for both types of
dimers).

It is worth mentioning that the molecular field model in which the
neighboring dimers mutually interact gave poor agreement with the
measured data. The possible explanation lies in the fact that at
low temperatures the antiferromagnetically coupled dimers are in
the singlet state, and their mutual interactions are therefore
unlikely.

The magnetic susceptibility analysis showed the presence of
magnetic dimerization in both DPPH1 and DPPH2, but the amounts of
entities participating and the strength of exchange couplings are
different for the two samples, in accord with their crystal
structures.


\section{Conclusions}
Crystal structures for two DPPH samples were solved: DPPH1,
crystallized from ether, and DPPH2, crystallized from \ce{CS2}.
The single-crystal X-ray diffraction analysis (and also IR
spectroscopy) showed that the single crystals of DPPH1 are solvent
free and those of DPPH2 contain one molecule of \ce{CS2} in the
unit cell. From the EPR measurement principal values of the
$g$-tensors at room (297~K) and low (10~K) temperatures were
obtained. Although the crystals of DPPH2 give a narrower
linewidth, the crystals of DPPH1, due to an almost insignificant
change of linewidth with decreasing temperature (from room
temperature to $T=10$~K) and a lower $g$-tensor anisotropy, prove
to be more suitable as the EPR probe. The magnetization study show
pairing into dimers with the antiferromagnetic exchange coupling
of $-17.4$~K for all molecules in DPPH1 and pairing into two kinds
of dimers (\emph{ca} 50-50\%) with the antiferromagnetic exchange
couplings of $-1.56$~K and $-83.9$~K, in DPPH2. The magnetization
results are in accordance with the crystal structures of the
compounds.

The results presented in this study contribute to better
understanding of the properties of DPPH, which is important
regarding the great significance of DPPH in the EPR spectroscopy.


\bigskip
\textbf{Acknowledgments}
\bigskip

D.\,\v{Z}ili\'{c} is grateful to D.\,Merunka for the useful
discussion. This research was supported by the Ministry of
Science, Education and Sports of the Republic of Croatia (projects
098-0982915-2939, 098-0982904-2946, 098-1191344-2943 and
119-1191458-1017).

\bibliographystyle{elsarticle-num}

\end{document}